\documentclass{jpsj2}
\usepackage{color}
\def\beq{\begin{equation}}
\def\eeq{\end{equation}}
\def\beqa{\begin{eqnarray}}
\def\eeqa{\end{eqnarray}}
\def\Nd{N_\mathrm{d}}

\title{
Mechanism of Slow Relaxation due to Screening Effect in a Frustrated System
}
\author{Shu \textsc{Tanaka}\thanks{E-mail address:
shu-t@issp.u-tokyo.ac.jp}
and
Seiji \textsc{Miyashita}$^{1,2}$\thanks{E-mail address:
miya@spin.phys.s.u-tokyo.ac.jp}
}
\inst{
Institute for Solid State Physics, University of Tokyo, 5-1-5 Kashiwanoha, Kashiwa-shi, Chiba 277-8581, Japan\\
$^{1}$ Department of Physics, University of Tokyo, 7-3-1 Hongo, Bunkyo-ku, Tokyo
113-0033, Japan\\
$^{2}$ CREST, JST, 4-1-8, Honcho Kawaguchi, Saitama 332-0012, Japan\\
}

\abst{
We study a slow relaxation process in a frustrated spin system in which
a type of screening effect due to a frustrated environment plays an
important role. This screening effect is attributed to the highly degenerate
configurations of the frustrated environment. This slow relaxation
is due to an entropy effect and is different from those due to the
energy barrier observed in systems such as random ferromagnets. In the
present system, even if there is no energy gap, the slow relaxation
still takes place. Thus, we call this phenomenon ``entropic slowing down''.
Here, we study the mechanism of entropic slowing down quantitatively in an Ising spin model with the so-called decorated bonds. 
The spins included in decorated bonds (decoration spins) cause a
peculiar density of states, which causes on entropy-induced screening effect.
We analytically estimate the time scale of the system that increases exponentially with the number of decoration spins.
We demonstrate the scaling of relaxation processes using the time scale at all temperatures including the critical point.
}
\kword{
slow relaxation,
Ising spin system,
frustration,
decorated bond system,
entropy effect
}
\begin{document}
\maketitle

\section{Introduction} 
The mechanisms of slow relaxation in strongly interacting systems have been studied over the past several decades. One of them is the critical slowing down. \cite{Kawasaki,Suzuki-Kubo,SM-Takano,Ito}
When the correlation of an order parameter develops near a critical point, the system shows a slow relaxation.
If the interactions of a system are nonuniform, as in the diluted
ferromagnetic model, the model shows a slow relaxation even in the
off-critical region. After quenching from a random state the order
develops locally, but the domain walls between them are pinned in
regions of relatively weak interaction. This pinning causes a very slow
relaxation that may be called a frozen state.\cite{Huse-Fisher,SM-Kawasaki,Takano-SM}. 
Furthermore, when the system has frustration, as in spin glass, the
frustration causes a random distribution of effective interactions and
the relaxation also becomes slow.\cite{Vincent,Nordblad,Bouchaud,Mezard,Fischer}
In a frustrated system, many competing configurations are degenerate,
and the system has a peculiar density of state that causes a temperature-dependent ordering structure\cite{Miyashita2,Tanaka1,Miyashita1}, and even reentrant phase transitions~\cite{Syozi,Kitatani,Tanaka1}.

Recently, it has been found that an extremely slow relaxation takes place
in a decorated lattice.\cite{Tanaka1} 
The static properties of the decoration lattice are described by an
effective coupling due to a decorated bond, and are the same as those
of a regular lattice that has a coupling constant, but the dynamics of
the systems are very different because the motion of decoration spins
prevents the smooth evolution of ordering. This screening effect comes from the distribution of degrees of freedom of decoration spins, and we call it ``entropic slowing down.''
This slow relaxation occurs even if there is no energy barrier.

The purpose of the present study is to clarify the feature of the
entropic slowing down, and to quantitatively analyze the time scale of the slowing down. 
Using the time scale, we demonstrate a scaling analysis of the relaxation processes obtained by a Monte Carlo simulation with the Glauber dynamics\cite{Glauber}.

Using the present slow-relaxation mechanism, we propose a new memory
system that we call the ``spin blackboard''. In such a system, we can memorize all configurations at low temperatures and erase them by increasing the temperature.

In \S\,2, we first introduce the model and present its equilibrium
and dynamical properties.
We present the probability distribution of the frustrated local structure in
\S\,3.
The effective time scale can be calculated analytically.
In \S\,4, we study the relaxation of magnetization on a square lattice
system by Monte Carlo simulation and perform the scaling plot using the
effective time obtained in \S\,3.
We provide a summary of this study in \S\,5.
In Appendix, we analyze the flip probability by another energy unit.

\section{Model and Equilibrium Properties}
\begin{figure}[h] 
\begin{center}
$$\begin{array}{cc}
\includegraphics[width=3cm]{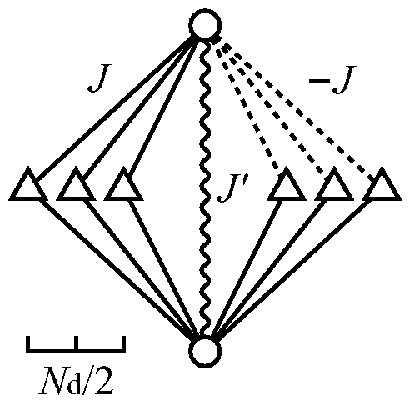}&
\includegraphics[width=3cm]{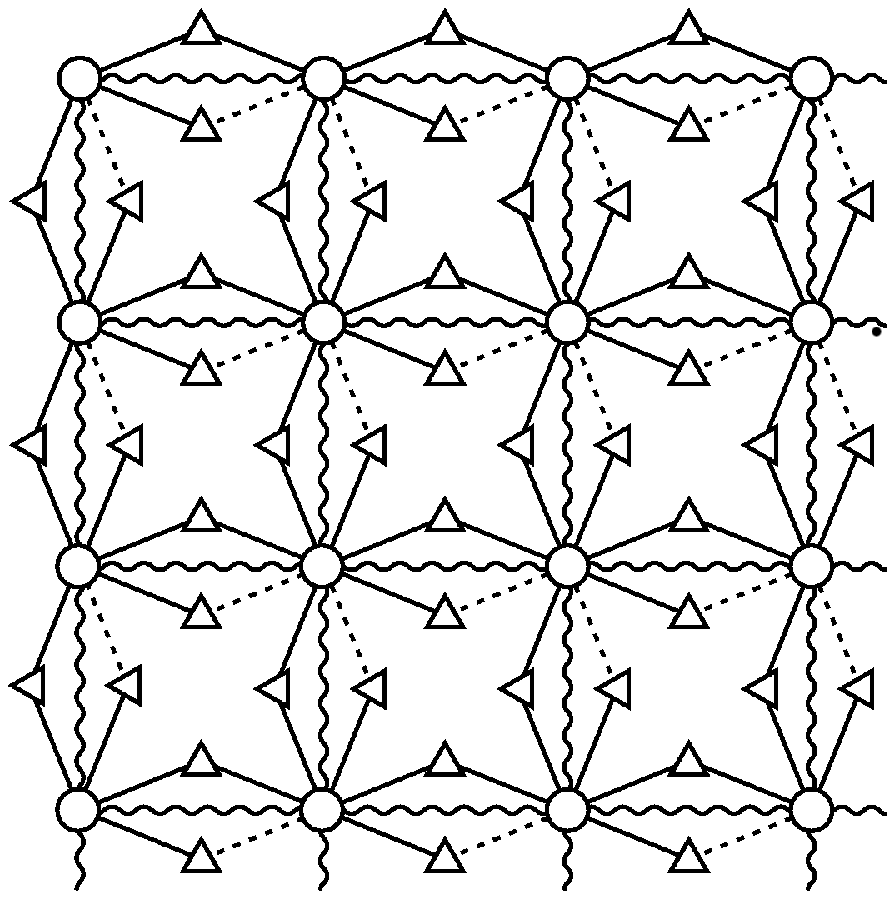}\\
({\rm a}) & ({\rm b})
\end{array}$$
\end{center}
\caption{(a) Frustrated decorated bond unit. The circles and triangles denote
the system and decoration spins, respectively.
The solid, dotted, and wavy lines denote the magnetic interactions of $J$, $-J$, and $J'$,
respectively.
(b) Two-dimensional square lattice with decorated bonds in the case of $\Nd=2$.
}
\label{Fig:model}
\end{figure}

We have studied the ordering process of a decorated bond system. A
decorated bond system consists of the so-called skeleton lattice,
consisting of ``system spins'' and the decorated bonds between them.  In
each decorated bond, the system spins are connected by a bunch of paths
that are mutually frustrated, as depicted in Fig.~\ref{Fig:model}(a).
To study the entropy effects on relaxation phenomena, we introduce the
two-dimensional square lattice Ising spin system with decorated bonds
depicted in Fig.~\ref{Fig:model}(b). In Fig.~\ref{Fig:model}(a), the frustrated local structure of the decorated bond unit is depicted. 
The circles denote ``system spins'' which form the square lattice as
depicted in
Fig.~\ref{Fig:model}(b). The triangles denote ``decoration spins''
$\{ s_i \}$, that are placed in the structure of a decorated bond.
There are $\Nd$ decoration spins between a nearest-neighbor pair of the system spins $\sigma_1$ and $\sigma_2$. 
We adopt the following structure of the decorated bond. We set half of
the paths in a decorated bond in a ferromagnetic way, {\it i.e.}, by two ferromagnetic bonds (solid lines) with a decoration spin between them.
The other half of the paths are set in an antiferromagnetic way with ferromagnetic and antiferromagnetic bonds (dotted lines) with a decoration spin between them.
The magnitudes of these bonds are set to be the same ($J$).
As we will show below, the contributions of the ferromagnetic and antiferromagnetic paths cancel out, and the effective interaction between the system spins due to the $\Nd$ paths is zero.

In order to study the phase transition, in addition, we set the extra
bond (a wavy line) that causes the interaction between the system spins
to be nonzero. We impose the periodic boundary condition on the lattice
depicted in Fig.~\ref{Fig:model}(b).

The Hamiltonian of a decorated bond depicted in Fig. \ref{Fig:model}(a) is
\begin{equation}
\mathcal{H} = -J' \sigma_1 \sigma_2
-J \sum_{i=1}^{\Nd/2} s_i \left( \sigma_1 + \sigma_2 \right)
-J \sum_{i=\Nd /2+1}^{\Nd} s_i \left( - \sigma_1 + \sigma_2 \right),
\end{equation}
where $\sigma_1$ and $\sigma_2$ denotes the system spins and $s_i$ ($i =
1,\cdots, \Nd$) denote the decoration spins.
The solid, dashed, and wavy lines in Fig.~\ref{Fig:model} denote the magnetic interactions of $J$, $-J$, and $J'$, respectively.
Hereafter, we take $J$ as the unit of energy.
The effective coupling $K_{\rm eff}$ between the system spins $\sigma_1$ and $\sigma_2$ at a temperature $T=1/\beta$ is defined by tracing out the decoration spins $\{ s_i \}$:
\begin{equation}
\sum_{\{s_i = \pm 1\}} \mathrm{e}^{-\beta\mathcal{H}}
= A \mathrm{e}^{-\beta \mathcal{H}_{\mathrm{eff}}}
= A \mathrm{e}^{-K_{\mathrm{eff}}\sigma_1 \sigma_2},
\end{equation}
where
\begin{eqnarray}
 A &=& \left( 4 \cosh 2\beta J\right)^{\Nd/2},\\
K_{\mathrm{eff}} &=& \beta J'.
\end{eqnarray}
Since the contributions of the left and right-half paths in
Fig.~\ref{Fig:model}(a) cancel out, the effective coupling of this system comes only from $J'$.
The relation between the correlation function
$\left\langle \sigma_1 \sigma_2 \right\rangle$ and the effective coupling
$K_{\mathrm{eff}}$ is given by
\begin{equation}
\left\langle \sigma_1 \sigma_2 \right\rangle = \tanh K_{\mathrm{eff}}.
\end{equation}
If $J' > 0 $, $\left\langle \sigma_1 \sigma_2 \right\rangle$ is positive.
If the effective coupling $K_{\mathrm{eff}}$ is larger than the critical value
$K_{\mathrm{c}}$, the system spins have a ferromagnetic long-range order in the equilibrium state. Here, we consider the case of $J' = 1$.

Next, we consider the square lattice system with decorated bonds depicted
in Fig.~\ref{Fig:model}(b).
We perform a single spin flip of the heat bath method of Monte Carlo simulation to consider the dynamical aspects in the system. We study a relaxation process of the system magnetization
\beq
M_{\rm s} = \frac{1}{N} \sum_i^{N} \sigma_i, 
\eeq
from an ordered state in which all the spins are aligned in the same direction.
Here, we set the temperature in the paramagnetic region $T=3$, {\it i.e.},
$K_{\rm eff}=1/3 < K_{\rm c} = \left[ \frac{1}{2}\log\left( 1+\sqrt{2}\right)
\right]^{-1}= \left( 2.27 \cdots \right)^{-1}$.
At this temperature, we expect a fast relaxation to $M_{\rm s}=0$ in the
regular lattice ($\Nd=0$). However, in the decoration bond system
($\Nd>1$), we expect the effect of the frustrated configuration although the thermodynamic properties, such as the correlation functions of the system spin, are the same in both systems.
 
\begin{figure}[t] 
\begin{center}
\includegraphics[width=5cm]{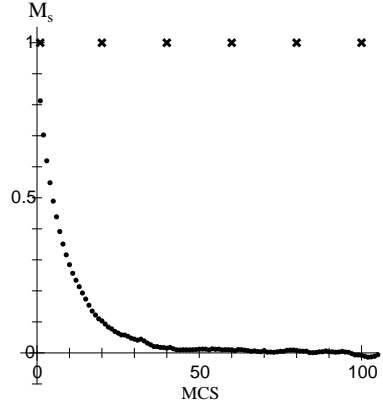}
\end{center}
\caption{
Relaxation process of the system magnetization on a square lattice
 system with decorated bonds, where the number of system spins is
 $N=20^2$, at $T=3$, which is above the critical temperature.
The circles indicate the relaxation process of the regular system ($\Nd = 0$)
and the crosses represent that of the decorated bond system of $\Nd = 200$.
}
\label{Graph:singledata}
\end{figure}

We performed a Monte Carlo simulation of the square lattice system
depicted in Fig.~\ref{Fig:model}(b) with $N =20^2$ system spins. In Fig.~\ref{Graph:singledata}, we compare the relaxations of $M_{\rm s}$
in the regular and decorated lattices. The data are obtained by taking
the average of one thousand samples.
The magnetization in the regular lattice relaxes with a relaxation time
$\tau_{\rm reg}\simeq 10$ Monte Carlo Step (MCS) we see no
relaxation in the decorated bond system in this time scale. 
In the next section we will focus on the microscopic mechanism of this slow relaxation in decorated bond systems.

\section{Probability Distribution of Frustrated Local Structure}

\begin{figure}[b] 
$$\begin{array}{cc}
\includegraphics[width=3cm]{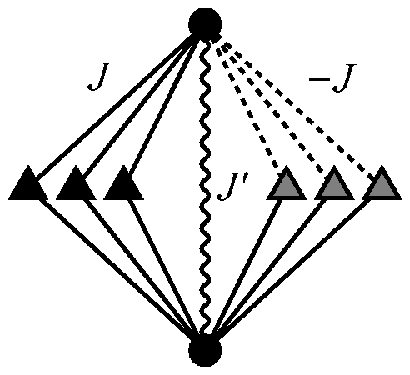}&
\includegraphics[width=3cm]{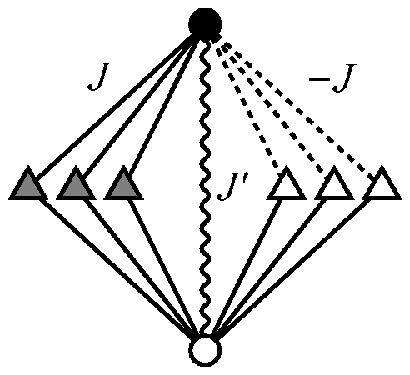}\\
({\rm a}) & ({\rm b}) \end{array}
$$
\caption{Black, white, and gray symbols denote the sites with $+$, $-$, and 0 internal fields, respectively.
(a) Parallel system spin case and 
(b) antiparallel system spin case.}
\label{Fig:ESDS-state}
\end{figure}

In the previous section, we showed an example of the entropic slowing down by Monte Carlo simulation. A large number of degenerate states cause such a slowing down.
In this section, we analyze this slow relaxation from the viewpoint of
the probability distribution of the decoration spins depicted in Fig.~\ref{Fig:model}(a).
We determine the number of states for a fixed local configuration of system spins. 
We depict ``the parallel state'' of system spins, {\it i.e.}, ($\sigma_1$,$\sigma_2$)$=$($+,+$), and ``the antiparallel state'', {\it i.e.},
($\sigma_1$,$\sigma_2$)$=$($+,-$) 
in Figs.~\ref{Fig:ESDS-state}(a) and (b), respectively, where the black, white, and gray symbols denote the sites with $+$, $-$, and 0 internal fields, respectively.
We denote the number of $+$ decoration spins in the ferromagnetic paths by $m$,
and that in the antiferromagnetic paths by $n$.
The energy of the parallel state is given by
\begin{equation}
\label{Eq:epp}
E_{++}^{\left( \Nd \right)} \left( m,n \right)
= \left( \Nd - 4m \right) J - J' .
\end{equation}
The energy of the parallel state does not depend on $n$, because the energy of each antiferromagnetic path does not depend on the states of decoration spins.
In the same way, the energy of the antiparallel state is given by
\begin{equation}
\label{Eq:epm}
E_{+-}^{\left( \Nd \right)} \left( m,n \right)
= \left( - \Nd + 4n \right) J + J',
\end{equation}
which is independent of $m$. Note that the following relations are satisfied:
\begin{eqnarray}
  \label{Eq:same-relation1}
  E_{++}^{\left( \Nd \right)} \left( m,n \right) =
  E_{--}^{\left( \Nd \right)} \left( \frac{\Nd}{2}-m,n \right), \\
  \label{Eq:same-relation2}
  E_{+-}^{\left( \Nd \right)} \left( m,n \right) =
  E_{-+}^{\left( \Nd \right)} \left( m,\frac{\Nd}{2}-n \right).
\end{eqnarray}
Because the maximum values of $m$ and $n$ are ${\Nd}/{2}$, the ground state energy of the parallel state is given by
\begin{equation}
E_{++}^{\left( \Nd \right)} \left( \frac{\Nd}{2},n \right) = -\Nd J - J'
\end{equation}
and the number of degeneracies is $2^{\Nd / 2}$.
Similarly, the minimum energy of the antiparallel state is given by
\begin{equation}
E_{+-}^{\left( \Nd \right)} \left( m,0 \right)
= -\Nd J + J' .
\end{equation}
The lowest energy of the antiparallel state ($n=0$) is higher than that of the parallel state by $2J'$.

We consider the probability distribution of the decoration spins at a temperature $T$.
First, we consider the parallel case. The probability of $m$ up spins in the $\Nd / 2$ ferromagnetic paths is given by
\begin{equation}
Q^{\left( \Nd \right)}_{++}\left( m \right) =
\frac{\mathrm{e}^{-\beta J \Nd}}{\left( 2\cosh 2\beta J \right)^{\Nd /2}}
\left(
\begin{array}{c}
\Nd /2 \\ m
\end{array}
\right)
\mathrm{e}^{4\beta J m},
\end{equation}
and the probability of $n$ up spins in the antiferromagnetic paths is given by
\begin{equation}
R^{\left( \Nd \right)}_{++}\left( n \right) =
\left(
\begin{array}{c}
\Nd /2 \\ n
\end{array}
\right)
\left( \frac{1}{2} \right)^{\Nd /2}.
\end{equation}

On the other hand, if the system is in the antiparallel case, the probability of $m$ up spins in the ferromagnetic paths is given by
\begin{equation}
\label{Eq:qpm}
Q^{\left( \Nd \right)}_{+-}\left( m \right) =
\left(
\begin{array}{c}
\Nd /2 \\ m
\end{array}
\right)
\left( \frac{1}{2} \right)^{\Nd /2},
\end{equation}
which is equal to $R^{\left( \Nd \right)}_{++}\left( m \right)$.
The probability of $n$ up spins in the antiferromagnetic paths is given by
\begin{equation}
\label{Eq:rpm}
R^{\left( \Nd \right)}_{+-}\left( n \right) =
\frac{\mathrm{e}^{\beta J \Nd}}{\left( 2\cosh 2\beta J \right)^{\Nd /2}}
\left(
\begin{array}{c}
\Nd /2 \\ n
\end{array}
\right)
\mathrm{e}^{-4\beta J n},
\end{equation}
which is equal to $Q_{--}^{\left( \Nd \right)} \left( m \right)$.
From the symmetry, the following relations hold:
\begin{eqnarray}
  Q^{\left( \Nd \right)}_{++}\left( m \right) &=& Q^{\left( \Nd \right)}_{--}\left( \frac{\Nd}{2}-m \right),\\
  R^{\left( \Nd \right)}_{++}\left( n \right) &=& R^{\left( \Nd \right)}_{--}\left( n \right),\\
  Q^{\left( \Nd \right)}_{+-}\left( m \right) &=& Q^{\left( \Nd \right)}_{-+}\left( m \right),\\
  R^{\left( \Nd \right)}_{+-}\left( n \right) &=& R^{\left( \Nd \right)}_{-+}\left( \frac{\Nd}{2}-n \right).
\end{eqnarray}
Because $R^{\left( \Nd \right)}_{++}(n)$ and $Q^{\left( \Nd \right)}_{+-}(m)$ are simple binary distributions,
they are independent of temperature.
The distributions become sharp when the number of decorated spins increases.
On the other hand, 
$Q^{\left( \Nd \right)}_{++}(m)$ and $R^{\left( \Nd \right)}_{+-}(n)$ depend on temperature.
At high temperatures, they are maximum at $\Nd /4$ owing to the
entropy effect. At low temperatures, they are maximum at nearly $\Nd /
2$ and $0$, respectively, because they are energetically favored states
for the distributions. 

\begin{figure}[h] 
$$\begin{array}{ccccc}
        \includegraphics[scale=0.5]{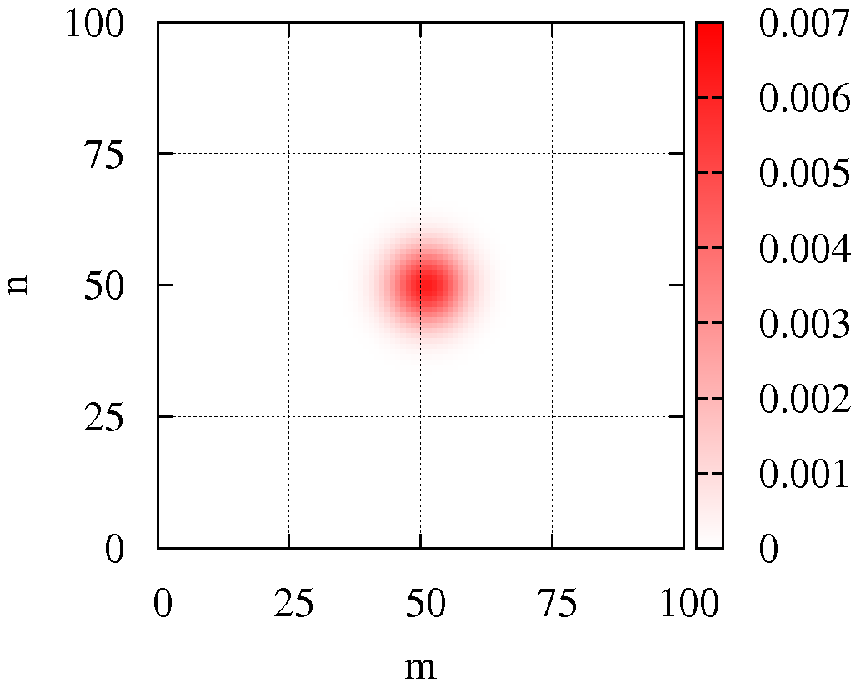}&\hspace{-50mm}&
        \includegraphics[scale=0.5]{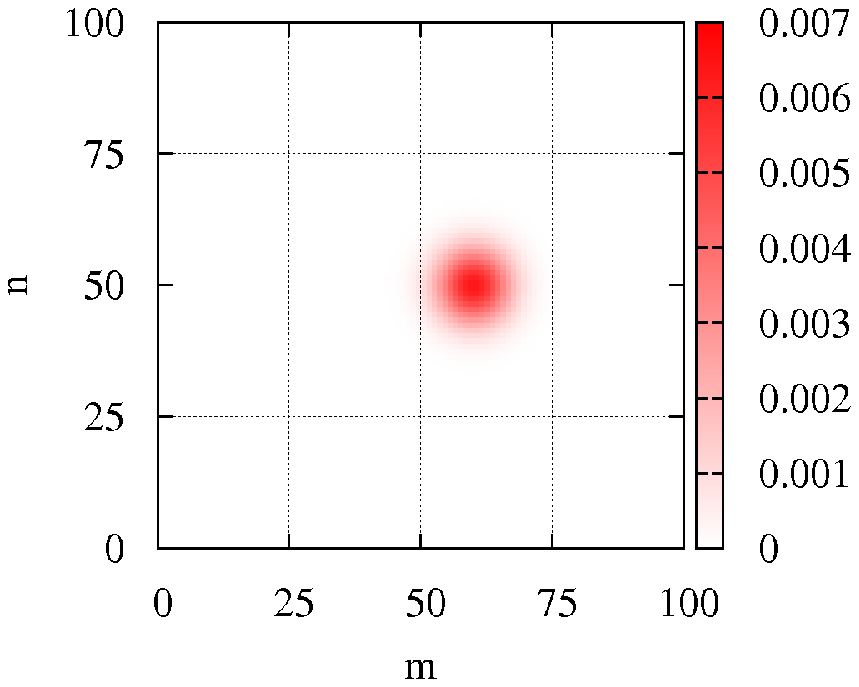}&\hspace{-50mm}&
	\includegraphics[scale=0.5]{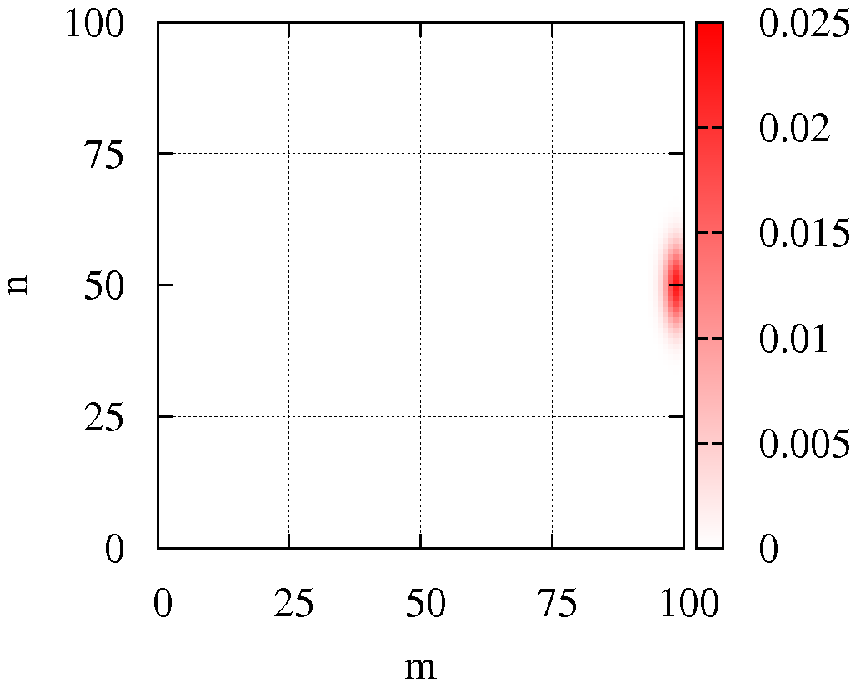}\\
        ({\rm a}) && ({\rm b}) && ({\rm c}) \\
        \includegraphics[scale=0.5]{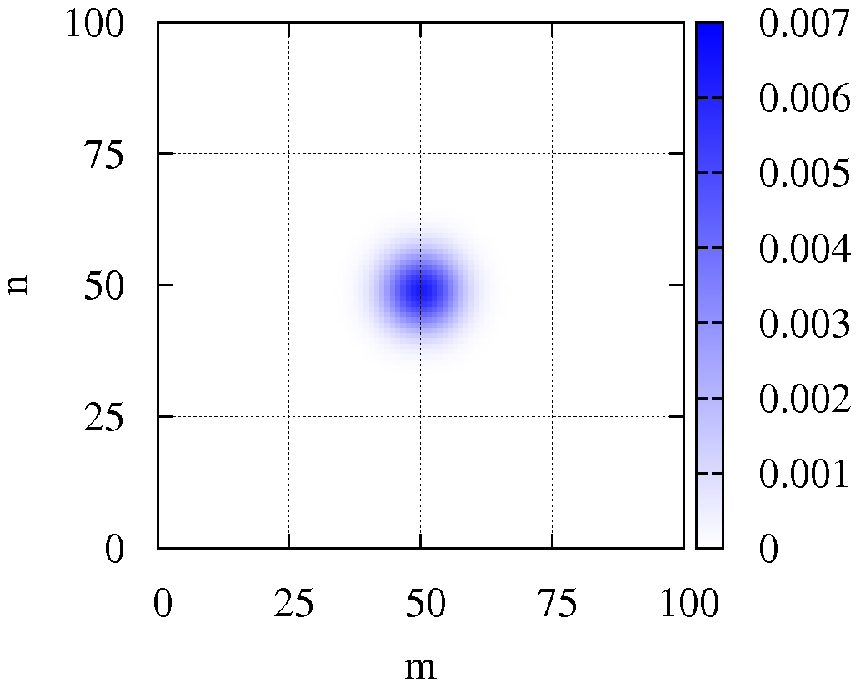}&\hspace{-50mm}&
        \includegraphics[scale=0.5]{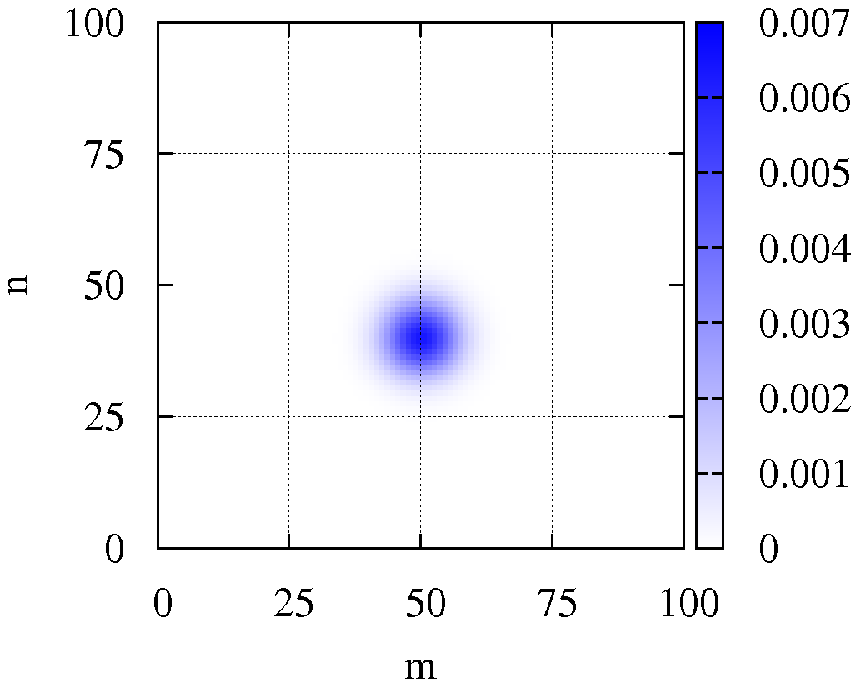}&\hspace{-50mm}&
	\includegraphics[scale=0.5]{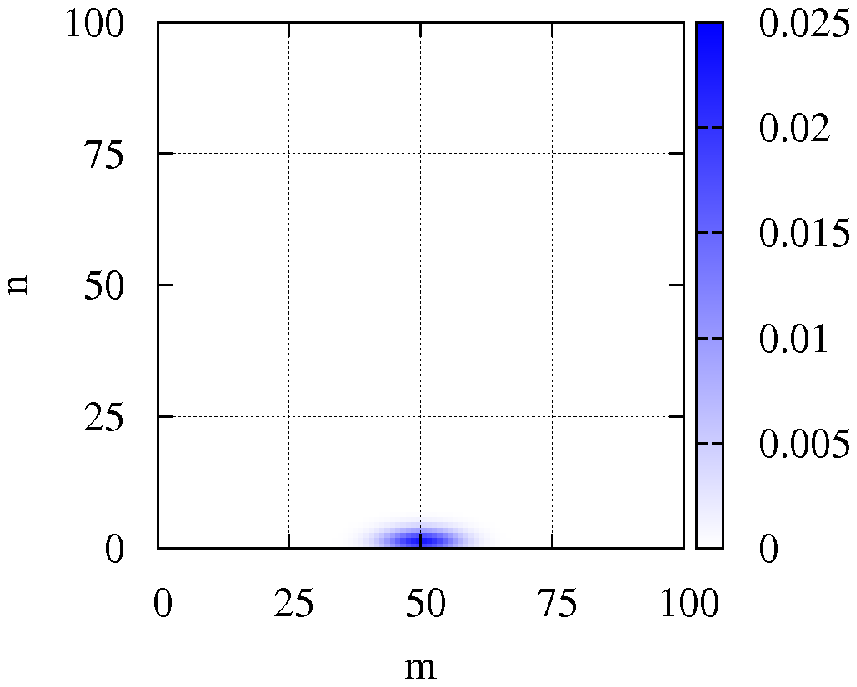}\\
        ({\rm d}) && ({\rm e}) && ({\rm f})  \\
        \includegraphics[scale=0.5]{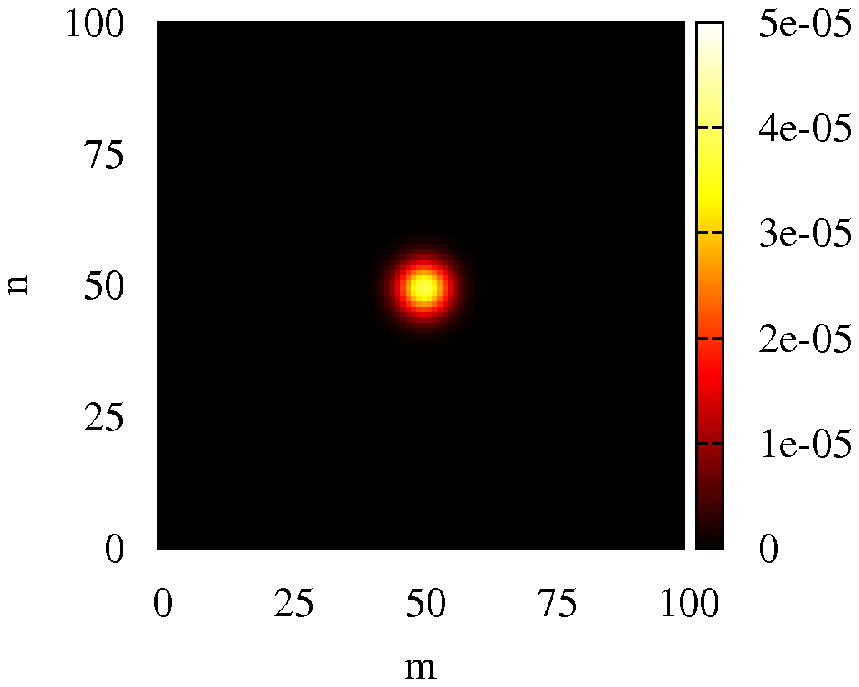}&\hspace{-50mm}&
        \includegraphics[scale=0.5]{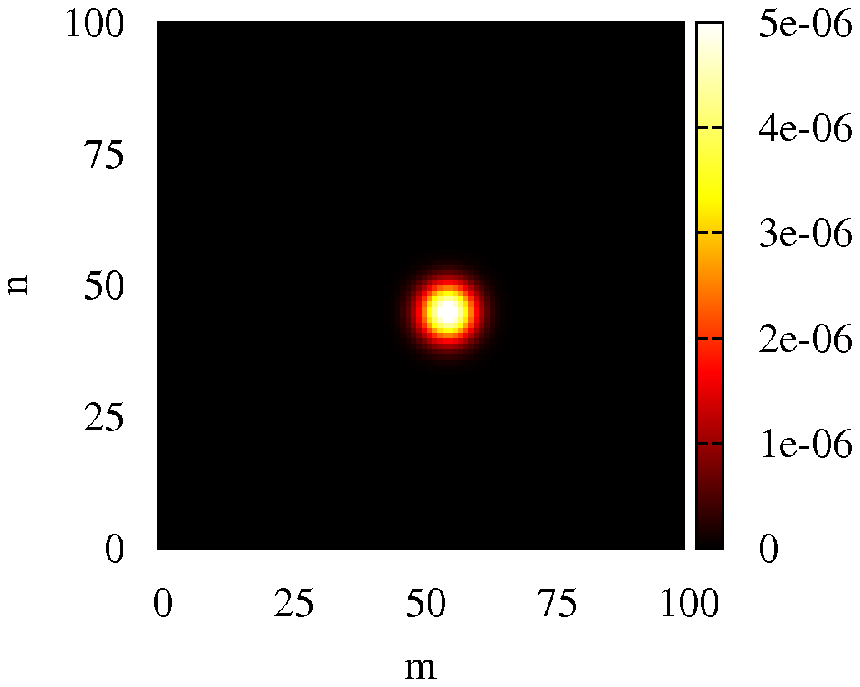}&\hspace{-50mm}&
	\includegraphics[scale=0.5]{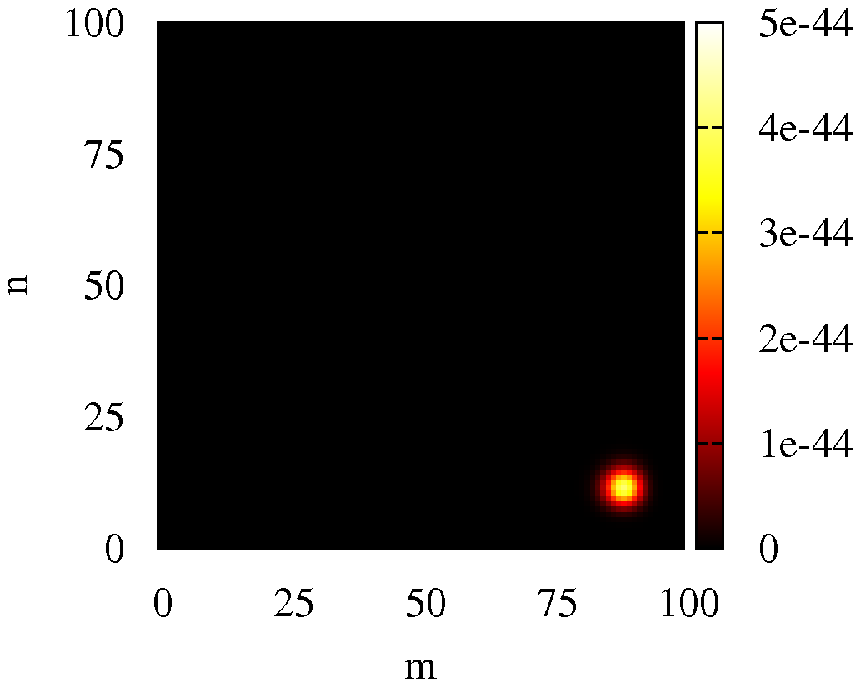}\\
        ({\rm g}) && ({\rm h}) && ({\rm i})  
\end{array}
$$
\caption{
(Color online) 
 We set the number of decoration spins $\Nd = 200$.
 (Top row) Probability distribution function of parallel
 state $P_{++}^{\left( \Nd \right)}(m,n)$ as a function of the numbers of
 $+$ decoration spins $m$ and $n$ in the cases of (a) $T=100$, (b)
 $T=10$, and (c) $T=1$.
 (Middle row) Probability distribution function of antiparallel
 state $P_{+-}^{\left( \Nd \right)}(m,n)$ as a function of the numbers of
 $+$ decoration spins $m$ and $n$ for (d) $T=100$, (e)
 $T=10$, and (f) $T=1$.
 (Bottom row) Overlap probability distribution function of parallel
 state $P_{++}^{\left( \Nd \right)}(m,n)$ and probability distribution
 function of antiparallel state $P_{+-}^{\left( \Nd \right)}(m,n)$ as a
 function of the numbers of $+$ decoration spins $m$ and $n$ in the case of (g) $T=100$, (h)
 $T=10$, and (i) $T=1$. Note that the ranges of
 values are not the same at all temperatures.}
\label{Graph:f-prob}
\end{figure}

The probability of the total distributions of the parallel and antiparallel states are given by
\begin{eqnarray}
\label{Eq:ppp}
P^{\left( \Nd \right)}_{++}\left( m,n \right) &=&
Q^{\left( \Nd \right)}_{++}\left( m \right)
R^{\left( \Nd \right)}_{++}\left( n \right), \\
\label{Eq:ppm}
P^{\left( \Nd \right)}_{+-}\left( m,n \right) &=&
Q^{\left( \Nd \right)}_{+-}\left( m \right)
R^{\left( \Nd \right)}_{+-}\left( n \right),
\end{eqnarray}
respectively.
Figure \ref{Graph:f-prob} shows color maps of the probability distributions 
$ P^{\left( \Nd \right)}_{++}\left( m,n \right)$, 
$ P^{\left( \Nd \right)}_{+-}\left( m,n \right)$, and their overlap
$ P^{\left( \Nd \right)}_{++}\left( m,n \right) 
P^{\left( \Nd \right)}_{+-}\left( m,n \right)$  as a
function of the numbers of $+$ decoration spins ($m$ and $n$) for $T=100$, $T=10$, and $T=1$ in the case of $\Nd =200$. 

The probability function is a binary distribution of $m$ and $n$ at high
temperatures, {\it e.g.}, $T=100$, where the overlap is large (left columns in Fig.~\ref{Graph:f-prob}).
As temperature decreases, the probability distributions split from
the center point and the overlap decreases (center columns of Fig.~\ref{Graph:f-prob}).
At low temperatures, {\it e.g.}, $T=1$, the peaks of the probability distributions
move to the edges because decoration spins with nonfrustrated
paths polarize, and the overlap becomes very small (right columns in Fig.~\ref{Graph:f-prob}). 
Therefore, the transition probability between the parallel and antiparallel state becomes very small at low temperatures.

Thus, if the system is initially in the antiparallel state, there is a
very small probability of transition to the parallel state, and the time
scale of the evolution of the configuration becomes very long. Therefore, the
system cannot reach the equilibrium state within a short time.

We denote the probabilities as 
$\mathcal{P}_{++}$ and $\mathcal{P}_{+-}$ for the parallel and antiparallel
configurations of the system spin obtained by tracing out the degree of freedom of decoration spins. 
They are given by
\begin{eqnarray}
\mathcal{P}^{\left( \Nd \right)}_{++} =
\sum_{m=0}^{\Nd /2} \sum_{n=0}^{\Nd /2} Q^{\left( \Nd \right)}_{++} \left( m \right)
R^{\left( \Nd \right)}_{++} \left( n \right) 
\frac{\mathrm{e}^{\beta J'}}{{\rm e}^{\beta J'} + {\rm e}^{-\beta J'}}
= {\mathrm{e}^{\beta J'} \over \mathrm{e}^{\beta J'} + {\mathrm{e}^{-\beta J'}}}, \\
\mathcal{P}^{\left( \Nd \right)}_{+-} =
\sum_{m=0}^{\Nd /2} \sum_{n=0}^{\Nd /2} Q^{\left( \Nd \right)}_{+-} \left( m \right)
R^{\left( \Nd \right)}_{+-} \left( n \right) 
\frac{\mathrm{e}^{-\beta J'}}{{\rm e}^{\beta J'} + {\rm e}^{-\beta J'}}
={\mathrm{e}^{-\beta J'}\over \mathrm{e}^{\beta J'}+ {\mathrm{e}^{-\beta J'}}},
\end{eqnarray}
and they agree with the thermal equilibrium probabilities.

\begin{figure}[t] 
\begin{center}
\includegraphics[width=7cm]{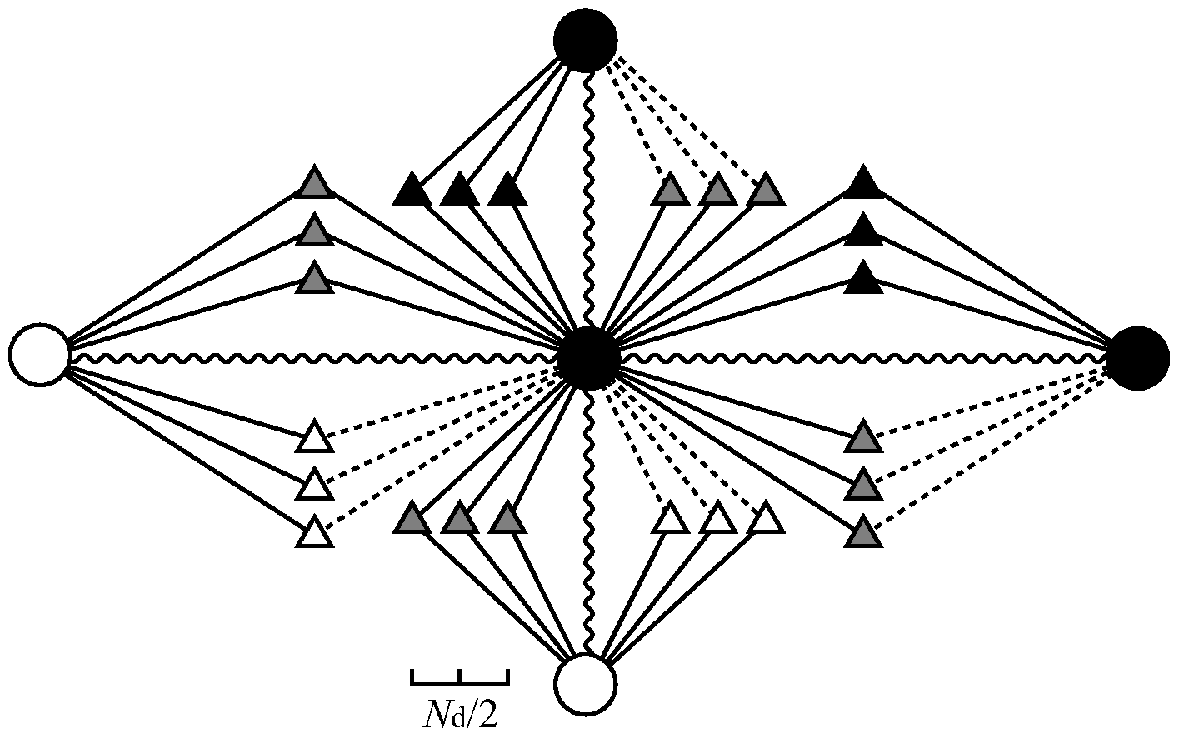}
\end{center}
\caption{Central system spin surrounded by two up system spins and
 two down system spins, which we call a ``free spin''.
The black, white, and gray symbols denote $+$, $-$, and disordered spins, respectively.
If decorated bonds are absent,
the flip probability of the center system spin is exactly $1/2$.
However, if decorated bonds exist,
the flip probability of the center system spin is less than $1/2$ owing to
 the entropy effect.
The number of decoration spins is $\Nd$ for each decorated bond.}
\label{Fig:freespin}
\end{figure}

To estimate the effective relaxation time of this system, we calculate the flip probability of the central spin in the configuration shown in
Fig.~\ref{Fig:freespin}.
This spin is surrounded by two up system spins and two down system spins.
Suppose we consider a regular spin system, {\it i.e.}, $\Nd = 0$. The
internal fields from four neighbor system spins cancel out, and the
energy difference between the up and down states of the central spin is zero. Thus, we call the central spin a ``free spin''. 
The transition probability of the center spin is $1/2$ in the Glauber dynamics\cite{Glauber}. 
However, as we will show below, in decorated bond systems $(\Nd
>0)$, the probability of the flip becomes smaller than $1/2$ owing to the
distribution of the surrounding decoration spins.

The internal field on the central spin in this configuration is
\begin{equation}
h\left(n_1, n_2, n_3, n_4 \right) = 2J \left( N_{\mathrm d} -n_1 + n_2 - n_3 - n_4
\right),
\end{equation}
where 
$n_1$ is the number of $+$ spins on the ferromagnetic paths in the antiparallel state of the system spins. 
Similarly, 
$n_2$ is that on the antiferromagnetic paths in the antiparallel state, 
$n_3$ is that on the ferromagnetic paths in the parallel state, 
and 
$n_4$ is that on the antiferromagnetic paths in the parallel state.

\begin{figure}[h] 
\begin{center}
\includegraphics[width=7cm]{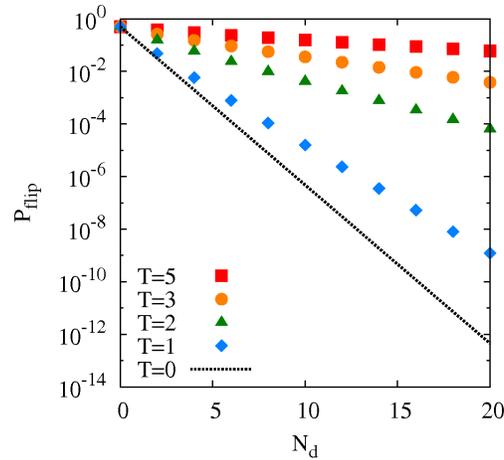}
\end{center}
\caption{
(Color online) Flip probability of the free spin $\mathcal{P}_{\mathrm{flip}}$ as a
function of the number of decoration spins for several temperatures.
The squares, circles, triangles, and diamonds denote the data for $T=5$,
 $3$, $2$, and $1$, respectively.
The low-temperature limit of the probability is shown by the dotted line.}
\label{Graph:pflip-finiteT}
\end{figure}

In the thermal bath transition probability, the flip probability of the center spin $\mathcal{P}_{\mathrm{flip}}$ is given by
\begin{eqnarray}
\nonumber
\mathcal{P}_{\mathrm{flip}} &=& \sum_{\left( n_1, n_2, n_3, n_4 \right)}
Q^{\left( 2\Nd \right)}_{+-}\left( n_1 \right) R^{\left( 2\Nd \right)}_{+-} \left( n_2 \right)
Q^{\left( 2\Nd \right)}_{++}\left( n_3 \right) R^{\left( 2\Nd \right)}_{++} \left( n_4 \right)\\
&&\times
\frac{1}{1+\exp\left[ -2 \beta h\left(n_1, n_2, n_3, n_4 \right)\right]}\\
\nonumber
&=& \frac{1}
{\left( 4\cosh 2\beta J \right)^{2\Nd}}
\sum_{\left( n_1, n_2, n_3, n_4 \right)}
\left(
\begin{array}{c}
\Nd \\ n_1
\end{array}
\right)
\left(
\begin{array}{c}
\Nd \\ n_2
\end{array}
\right)
\left(
\begin{array}{c}
\Nd \\ n_3
\end{array}
\right)
\left(
\begin{array}{c}
\Nd \\ n_4
\end{array}
\right)\\
\label{Eq:pflip-exact}
&&\times
\frac{\exp\left[ 4\beta J \left(n_3 - n_2 \right)\right]}{1+\exp\left[ -4 \beta
J \left( \Nd - n_1 + n_2 - n_3 - n_4 \right)\right]}.
\end{eqnarray}
%
Figure \ref{Graph:pflip-finiteT} shows the dependence of the flip probability of the free spin $\mathcal{P}_{\mathrm{flip}}$ as a function of $\Nd$ at $T=5$, $3$, $2$, $1$, and $0$.
Here, we find that the stabilization effect is significant when $\Nd$ is
large at low temperatures.

We define the time scale of the evolution by 
\beq
\tau_{\mathrm{eff}}={\mathcal{P}_{\mathrm{flip}}}^{-1}.
\label{tau}
\eeq
It increases rapidly with $\Nd$.
Note that the low-temperature limit of $\tau_{\rm eff}$ is
finite, because the slowing down is caused by the entropy effect, not by
energy. Namely, 
\beq
\lim_{T \to 0} \mathcal{P}_{\rm flip} =  \frac{1}{2} \left( \frac{1}{4} \right)^{\Nd}.
\eeq
This limit is shown by the dotted line in Fig.~\ref{Graph:pflip-finiteT}.

\section{Monte Carlo Simulation}

\begin{figure}[h]
\begin{center}
\includegraphics[width=7cm]{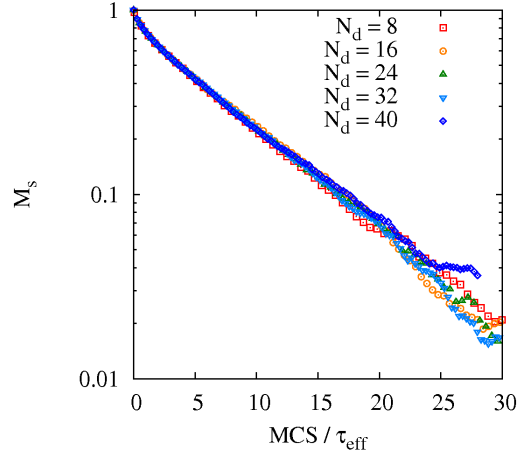}
\end{center}
\caption{
(Color online) Semilog plot of the relaxation of the system
 magnetization at $T=3$ which is above the critical temperature.
The squares, circles, triangles, inverted triangles and diamonds denote the case of $\Nd = 8$, $16$, $24$, $32$ and $40$,
 respectively. 
 Relaxation curves obey the exponential function: $M(t) = 0.86 {\rm e}^{-t/7.31}$.
}
\label{Graph:magscale}
\end{figure}

In the previous section, we calculated the effective time of a flipping spin.
We study the relaxation process of system magnetization on the
square lattice system shown in Fig.~\ref{Fig:model}(b) at $T=3$,
which is above the critical temperature in this section. 
We set $N=20^2$ as the number of system spins.
Data are obtained by taking the average over one thousand samples in cases of $\Nd=8$, $16$, $24$, $32$, and $40$.
As $\Nd$ increases, the relaxation of system magnetization becomes slow.
We scale these relaxation processes by the time scale
$\tau_{\mathrm{eff}} \left( \Nd \right)$ (Table \ref{Tab:teff})
estimated using eq.~(\ref{tau}), and plot them in
Fig.~\ref{Graph:magscale}.
\begin{table}[b]
\begin{center}

\begin{tabular}[t]{|c|c|c|c|c|c|}
\hline
 $\Nd$ & $8$ & $16$ & $24$ & $32$ & $40$ \\
\hline
 $\tau_{\mathrm{eff}}$ & $1.03037 \times 10^2$ & $3.01700 \times 10^3$ &
	     $7.96926 \times 10^4$ & $2.00850 \times 10^6$ & $4.92665
		     \times 10^7$\\
\hline
\end{tabular}
(a) $T=2$

\begin{tabular}[t]{|c|c|c|c|c|c|}
\hline
 $\Nd$ & $8$ & $16$ & $24$ & $32$ & $40$ \\
\hline
 $\tau_{\mathrm{eff}}$ & $5.43835 \times 10$ & $8.96284 \times 10^2$ &
	     $1.34004 \times 10^4$ & $1.91522 \times 10^5$ & $2.66641
		     \times 10^6$\\
\hline
\end{tabular}
(b) $T=T_{\rm c}$

\begin{tabular}[t]{|c|c|c|c|c|c|}
\hline
 $\Nd$ & $8$ & $16$ & $24$ & $32$ & $40$ \\
\hline
 $\tau_{\mathrm{eff}}$ & $1.77064 \times 10$ & $1.08986 \times 10^2$ &
	     $6.17373 \times 10^2$ & $3.36029 \times 10^3$ & $1.78598
		     \times 10^4$ \\
\hline
\end{tabular}
(c) $T=3$

\caption{Effective relaxation times of the free spin system
 $\tau_{{\rm eff}}$ at $T=2$, $T=T_{\rm c}$, and $T=3$ as a
function of the number of decoration spins $\Nd$.}
\label{Tab:teff}
\end{center}
\end{table}
All the relaxations of system magnetization converge in a scaling
function, which indicates that the analysis in the previous section works well.
In the case of $\Nd=200$ shown in Fig.~\ref{Graph:singledata},
effective relaxation time is estimated to be $\tau_{{\rm eff}} \sim
10^{20}$ MCS.
Therefore, we see no relaxation within $100$ MCS.

\begin{figure}[h] 
\begin{center}
\includegraphics[width=7cm]{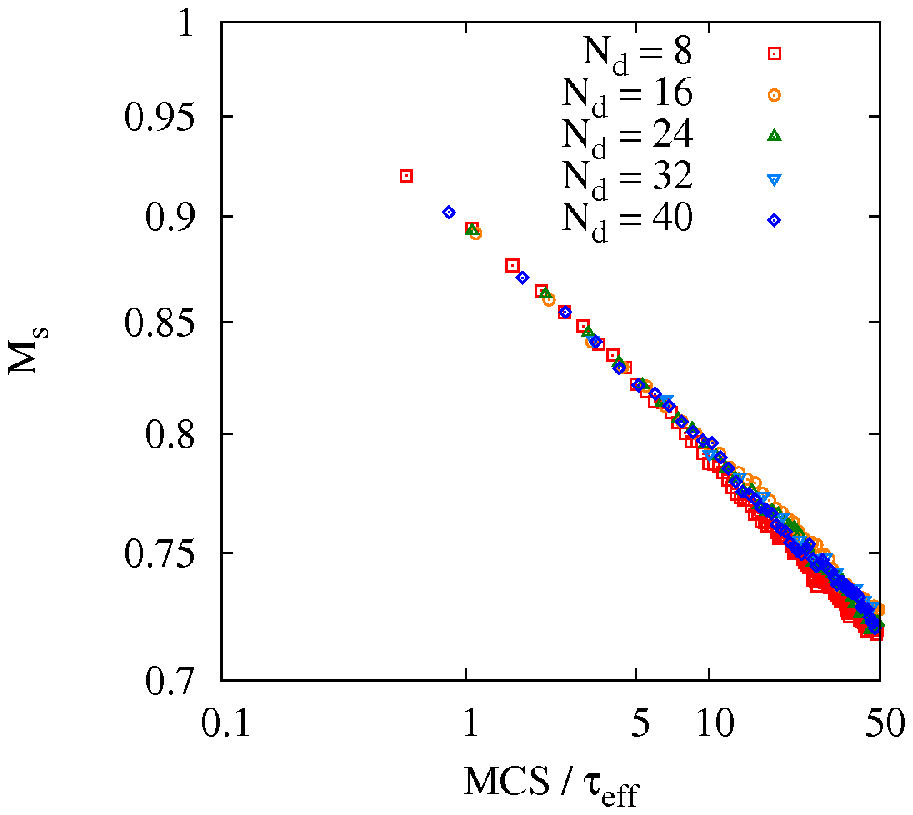}
\includegraphics[width=7cm]{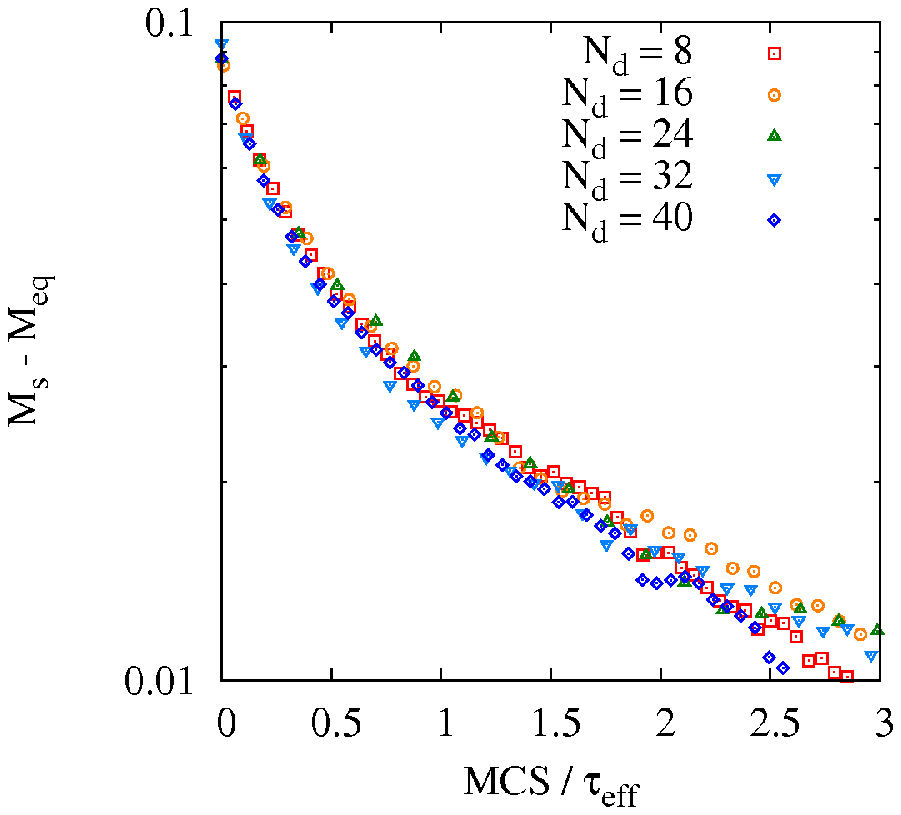}
\end{center}
\caption{
(Color online) Double logarithmic plot of the relaxation of the system
 magnetization at $T=T_{\rm c}$ (left panel) and semilog plot of the
 difference between system magnetization and the equilibrium value $M(t)
 - M_{\rm eq}$ at $T=2$, which is below
 the critical temperature (right panel).
 At $T=2$, the equilibrium value of magnetization $M_{\rm eq} = 0.911$.
 The symbols are the same as those in Fig.~\ref{Graph:magscale}.
 Magnetization curves obey the power function $M(t) = 0.90 t^{-0.057}$ at
 the critical point (left panel) and the exponential function $M(t) =
 0.061 {\rm e}^{-t/1.52} + 0.911$ at $T=2$ (right panel).
}
\label{Graph:magscaletct2}
\end{figure}

Next, we consider the slowing down at the critical temperature and below the critical temperature.
In the present model, the critical temperature is
$T_{\rm c} = \frac{1}{2} \log ( 1 + \sqrt{2}) J'$ because the effective coupling is given by $K_{\rm eff}=\beta J'$.
The relaxation curves at the critical temperature obey a power law decay,
while those below the critical temperature obey an exponential decay
toward spontaneous magnetization. In both cases, we find that the scaling works very well, as shown in  Fig.~\ref{Graph:magscaletct2}.
Therefore, we conclude that this mechanism of slowing down works at all temperatures.
Both above and below the critical temperature, the relaxation curves can
be fitted to the exponential function such as $M_{\rm s} (t) = A {\rm
e}^{-t/\tau} + M_{\rm eq}$.
In the case of $T=3$, we obtain the following parameters: $A=0.859$ and
$\tau = 7.31$ at $M_{\rm eq} = 0$.
In the same way, we obtain the following parameters in the case of
$T=2$: $A=0.0609$ and $\tau=1.52$ at $M_{\rm eq}
= 0.911$.
On the other hand, at the critical point, the relaxation curve obeys the
power law such as $M_{\rm s} (t) = At^{-b}$.
We obtain the following parameters: $A=0.90$, $b=0.0569$.

This slowing down mechanism holds the effect of the system spins as well
as that in the case of regular system ({\it i.e.}, with no decoration spins),
because the slowing down behavior is just a local effect.

If we set $J'=0$, then the system is in the complete paramagnetic state,
and all the spin
configurations of system spins have the same energy. However, because of the
entropic slowing down studied above, the system can maintain any
configuration for a long time. Thus, we may use such a system as a new
type of memory system, which we would call a spin blackboard. If we
increase the temperature, we can erase the information of spin configuration.

\section{Conclusion}

We studied the microscopic mechanism of the increase in the relaxation time of the decorated bond system.
We call this slow relaxation phenomenon ``entropic slowing down''.
The origin of the entropic slowing down is the large number of degenerate configurations due to frustration and not to the energy gap between metastable and stable states.
If we set an additional interaction such as $J'$, the ground state is an
ordered state. However, the system cannot reach the ground state
because of the freezing effect when it is cooled from a high-temperature disordered state.
The system is trapped in some random configuration in a temperature
region on the order of $J$, and no change occurs at low temperatures.
Even if there is no additional interaction, this entropic slowing down also appears.
We estimated effective relaxation time as a function of temperature and
the number of decoration spins analytically.
We also studied the relaxation process by a Monte Carlo simulation by the heat bath method.
The scaling analysis was successfully performed at all temperatures including the critical temperature.

Originally, the above model with decoration spins has been proposed to
explain the slowing down in the system where reentrant phase transition occurs\cite{Tanaka1}.
We expect that this entropic slowing down also appears in more complicated systems such as spin glass.
There, some parts of the random spin system are less frustrated where the spins are strongly correlated, while other parts are highly frustrated where the spins remain disordered.
We can interpret the former parts by ``system spins'' and the latter by ``decoration spins''.
In this course-grained picture, the spin order can be formed in the network of the former parts.

Moreover, we proposed a new type of memory system, {\it i.e.} the spin
blackboard in which any spin configuration can be stored owing to the entropy effect.
If we can prepare a ``micelle'' type lattice in which a system spin is
surrounded by many decoration spins, a spin blackboard device can be made.
We expect that some examples of entropic slowing down will be realized
and the spin blackboard behavior will be demonstrated experimentally.

The thermal annealing method\cite{Kirkpatrick,Kirkpatrick-2} is adopted
in many problems: it is a very efficient method of obtaining a stable state.
However, the thermal annealing method would not be efficient for obtaining the ground state in systems with entropic slowing down.
A number of researchers have studied the quantum annealing
method\cite{Finnila,Kadowaki,Suzuki,Das,Santoro,Morita,Tanaka2,Das-2,Matsuda,Tanaka3,Tanaka4,Tanaka5}
of obtaining the ground state of random systems as a substitute for thermal annealing.
We have found that the quantum annealing method is successful in determining the ground state of this entropic slowing down system. 
The mechanism of this method will be reported elsewhere.

\section*{Acknowledgments}
The authors would like to express their thanks to
Masaki Hirano, Naomichi Hatano and Eric Vincent for helpful
discussions. ST also thanks Atsushi Kamimura for the critical reading of this manuscript.
This work was partially supported by Research on Priority Areas
``Physics of new quantum phases in superclean materials'' (Grant No.
17071011) from MEXT,
and also by the Next Generation Super Computer Project, Nanoscience
Program from MEXT,
The authors also thank the Supercomputer Center, Institute for Solid State
Physics, University of Tokyo for the use of its facilities.

\appendix

\section{Another Energy Scale: $N_{\rm d}J$}

\begin{figure}[h]
 \begin{center}
  \includegraphics[width=7cm]{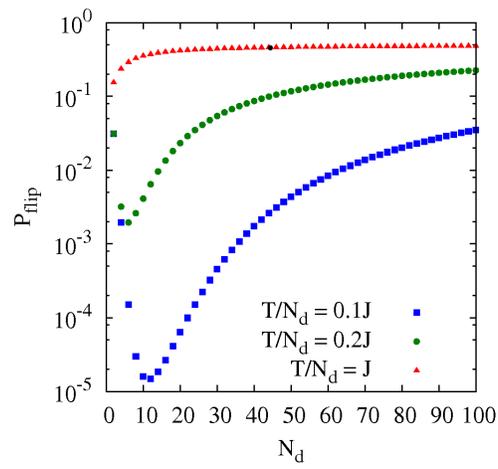}
  \caption{
  (Color online) Flip probability of the free spin
  $\mathcal{\tilde{P}}_{\rm flip}$ as a function of the number of
  decoration spins at several temperatures.
  The squares, circles, and triangles denote the data for $T/\Nd = 0.1J,
  0.2J$, and $J$, respectively.
  }
  \label{Fig:ndj-unit}
 \end{center}
\end{figure}

In this paper, we set the energy unit to be $J$.
In this appendix, we analyze  the flip probability $\mathcal{P}_{\rm
flip}$ regarding $N_{\rm d}J$ as the energy unit because the system
spins are surrounded $N_{\rm d}$ decoration spins.
In Fig. \ref{Fig:ndj-unit} we depict the flip probability of the free spin as
a function of the number of decoration spins for the reduced
temperature $T/N_{\rm d} = 0.1J$, $0.2J$, and $J$.
At low reduced temperatures, a nonmonotonic behavior appears, which can
be understood as follows.
The molecular field on a free spin from the outside decoration spins
is given by $-J N_{\rm d} \tanh \left( 2 \beta J\right)$.
When $\alpha = T/\left( N_{\rm d} J \right)$ is small, $N_{\rm
d}$ $-J N_{\rm d} \tanh \left( 2\beta J \right) \simeq J N_{\rm d}$ is
small and
thus molecular field increases with $N_{\rm d}$, which causes
a decrease in $P_{\rm flip}$.
On the other hand, when $\alpha = T/\left( N_{\rm d} J \right)$ is
large, $-J N_{\rm d} \tanh \left( 2\beta J \right) \sim = - 2J /\alpha$,
which does not depend on $N_{\rm d}$.
For a given $T/N_{\rm d}$, the temperature $T$ increases as $N_{\rm d}$
increases and thus $P_{\rm flip}$ approaches to its high temperature
limit $P_{\rm flip} = 0.5$.

\end{document}